# Implementation of Bounded Diffusion Impedance in a Model PyEIS to Correctly Simulate Flow Gradient on Channel-electrode in Microfluidics

Rassen Boukraa, Claire Poujouly, Pedro Gonzalez Losada, and Jean Gamby[z,*]
Université Paris-Saclay, CNRS, Centre de Nanosciences et de Nanotechnologies, 91120, Palaiseau, France
[z] E-mail: jean.gamby@universite-paris-saclay.fr
* ECS member

## Abstract
Here, we propose to test the limits of the bounded diffusion impedance model by using electrochemical impedance spectroscopy (EIS) measurements in a microfluidic device architecture bearing a set of microelectrodes integrated in microchannel. This EIS Nyquist plot behaviour was currently modelled with the Randles equivalent circuit. In this work, one of the main advantages for using PyEIS (a Python-based Electrochemical Impedance Spectroscopy) software is being able to modify and correct the models with a simple implementation in the source code freely accessible with published data [K. B. Knudsen, *ECS Meeting Abstracts*, MA2019-01, 1937 (2019)]. The obtained results after implementing the diffusion convective equation in the PyEIS code permits to adapt each function to be able to plot, simulate and fit a specific modified Randles for channel-electrode. The modified Randles was used for preliminary tests for optimization of the parameters of flow in the microfluidic channel by comparing the diffusion layer thickness as a function with the volumetric flow.

## 1. Introduction

The mutual use of a band microelectrode and a microfluidic channel named a channel-microelectrode presents two mains advantages. First, the obtention of a fast steady state current response ( ~ 1 ms) due to a thinner diffusion layer thickness compared to the convection layer. Second, it is an efficient strategy to drastically reduce the sample volume, to improve the system dynamics with easy manipulation, and to increase the sensor analytical performances. [1],[2],[3]. According to the model presented by Horny et. al. [4] using cyclic voltammetry (CV), the higher the flow rate is, the thinner is the diffusion layer thickness. A higher current density is moreover expected at the electrode when increasing the flow rate. It is however mandatory to keep a low enough value for the flux to stay in the limit of the pre-established model. Therefore, it relies on the dimensionless Péclet number (Pe) to quantify the convective flux versus diffusive flux ratio. [5] For Pe >> 1 close to the electrode, the convective effect dominates against the diffusive transport. So that, the flow rate is higher and the formation of a small diffusion layer is ensured. It often cannot match with experimental measurements because from a theoretical point of view, the expression of the diffusion impedance will be more affected by the wall velocity gradient fluctuations of the turbulent flow into the microchannel, and from a practical point of view, too much samples are circulated and wasted when high working flow rate is used. From these considerations, the previous work by Horny et al [4] has shown that the maximum flow rate would be attained around 1 µL/s taking into account.

Here, we proposed to test the limits of the model by using electrochemical impedance spectroscopy (EIS) in the same device architecture as in reference [4]. This EIS Nyquist plot behaviour was currently modelled with the Randles equivalent circuit [6]. To this work, one of the main advantages for using PyEIS (a Python-based Electrochemical Impedance Spectroscopy) software is being able to modify the pre-existing models with a simple implementation in the source code freely accessible on the website of the author Kristian B. Knudsen [7].

## 2. Experimental
### 2.1 Microfluidic chip and microelectrode Fabrication

As illustrated in **Figure 1** the microfluidic chip was composed of a set gold working microelectrode (WE) (30 µm x 300 µm) and platinum counter electrode (CE) (2 mm x 300 µm) patterned by photolithography and lift-off processes on glass wafers. Second, the chip was closed off with PDMS (ratio 1:10), patterned with a SU-8 mold and sealed by $O_2$ plasma bonding. The detailed protocol for WE gold electrode microfabrication is the same as described in Horny et al [4].

### 2.2 Measurement protocols
Electrochemical impedance spectroscopy (EIS) experiments were performed in an equimolar solution of 3 mM $[Fe_{(III)}(CN)_6]^{3-}/[Fe_{(II)}(CN)_6]^{4-}$, with flow rates from 0.01 to 1 µL/s. The electrochemical characterizations were done in a two electrodes setup configuration [3],[4] at equilibrium potential 0V with 10 mV excitation signal between 1MHz and 100 mHz.

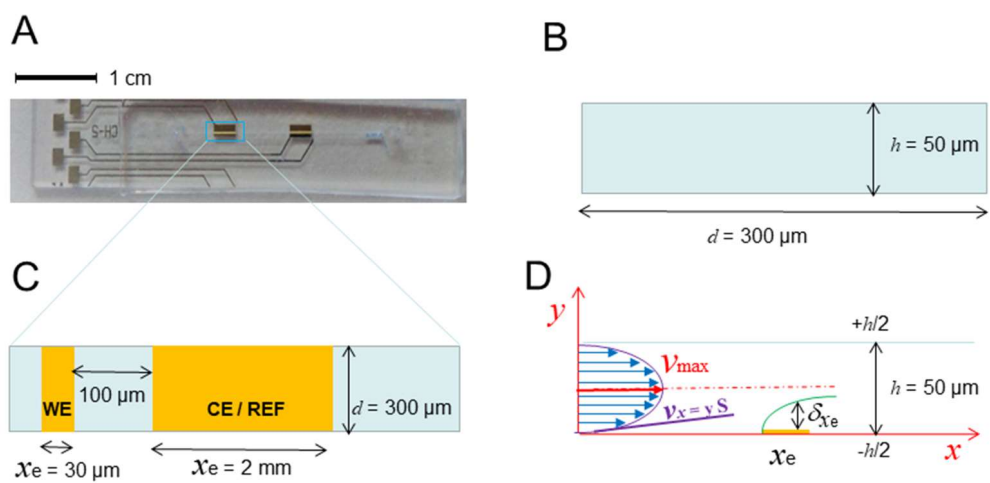

**Figure 1. A.** Global picture of the chip comprising two pairs of two-microelectrode networks. **B.** Schematic cross-section view for channel fluidic configuration where $h$ and $d$ represent the channel height and width, respectively. **C.** Schematic top view for microchannel electrode configuration where WE stands for working electrode (width, $d = 300$ µm, length $x_e = 30$ µm), CE stands for counter-electrode (d = 300 µm, $x_e$ = 2 mm). **D.** Schematic view of 2-D channel electrode with developed Poiseuille flow (high aspect $d/h$ ratio) with the steady-state diffusion-convection equation. $v_{max}$; $v_x$, $S$ and $\delta_{xe}$ represent the maximum velocity of the developed parabolic flow, the linear approximation of the flow velocity, the wall velocity gradient, and the local diffusion layer thickness.

## 3. Results

In the Nyquist plot representation of a traditional Randles Cell (**Scheme 1A**), the first loop is representative of the charge transfer resistance, $R_{ct}$, in parallel with a double layer capacitance, $C_{dl}$, or a constant phase element (CPE) capacitance, $Q_{dl}$, in order to take into account the non-ideal behaviour of the double layer capacitance. The second part of the diagram is a straight line representing the Warburg Impedance, $Z_W$, assuming the semi-infinite diffusion layer, as follows

$$Z_W = \sigma \cdot (\omega)^{-\frac{1}{2}} \cdot (1-j) \qquad (1)$$

in which $\omega$ is the angular frequency, and $\sigma$ represents the Warburg coefficient.

However, when mass transport in controlled by convection using rotating disk electrode (RDE) or channel-microelectrode as in microfluidics, it is obvious that the diffusion impedance layer thickness is said to be finite diffusion layer, as follows

$$Z_W = \sigma(\omega)^{-\frac{1}{2}}(1-j)\tanh(\delta \left(\frac{j\omega}{D}\right)^{\frac{1}{2}}) \qquad (3)$$

where $\delta$ is the Nernst diffusion layer thickness, and $D$ the average diffusion coefficient of the redox tracer.

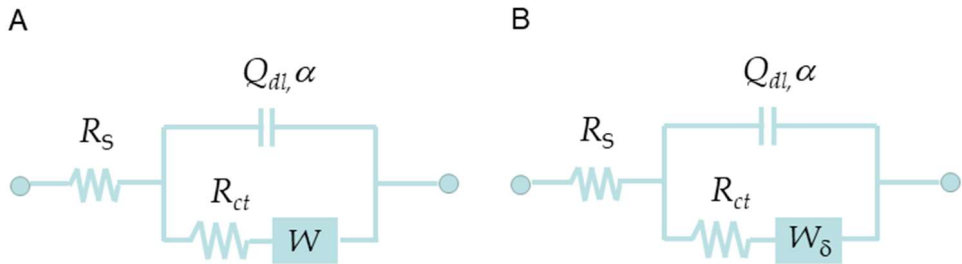

**Scheme 1. A**. Traditional Randles Cell with Warburg ($W$) impedance diffusion model. **B**. Implementation of bounded diffusion ($W_\delta$) impedance model in PyEIS to correctly simulate the thickness of the Nernst layer under channel hydrodynamics.

In order to implement this new reference model in the PyEIS, we have to determinate its equivalent impedance equation: $Z_{eq} = R_s + ((R_{ct} + Z_W)//Z_Q)$

Which give after development $Z_{eq} = R_s + 1/(\frac{1}{Z_Q} + \frac{1}{R_{ct}+Z_W})$

$$Z_{eq} = R_s + \frac{R_{ct}+Z_W}{1+C_{dl}\cdot(j\omega)^\alpha(R_{ct}+Z_W)} \qquad (4)$$

For $\omega \to 0$, the first order Taylor expansion of $Z_W$ gives

$$Z_W(0) = R_{W0} = \sigma(\omega)^{-\frac{1}{2}}(1-j)\delta\left(\frac{j\omega}{D}\right)^{\frac{1}{2}} = \sigma(1-j)\delta\left(\frac{j}{D}\right)^{\frac{1}{2}}, \text{ with } \sqrt{j} = \frac{1}{\sqrt{2}}(1+j) \text{ leads to}$$

$$R_{W0} = \sigma\frac{\delta}{\sqrt{D}}\cdot\frac{1}{\sqrt{2}}(1+j)(1-j)$$

and finally,

$R_{W0} = \sqrt{2\tau_D}\,\sigma$, where $\tau_D = \frac{\delta^2}{D}$, represents the time constant characteristic of the diffusion phenomenon.

We can rewrite the expression of the Warburg Impedance as a function of $R_{W0}$ and $\tau_D$ as follow,

$$Z_W = R_{W0}(1-j)\frac{\tanh(\sqrt{j\omega\tau_D})}{\sqrt{2\omega\tau_D}} \quad (5)$$

By analyzing the curves and the equation (4), we can guess approximate value for $R_s$ for $\omega \to +\infty$, at the first intersection between the curve and the x-axis. $R_{ct}$ can also be guessed by considering the minima reached between the two loops.
For $\omega \to 0$, the equation (4) gives $Z_{eq}(0) = R_s + R_{ct} + R_{W0}$. With determining first $R_s$ and $R_{ct}$, $R_{W0}$ it is then quite easy to approximate at the second intersection between the curve and the x-axis. Regarding this model, we relied on these first approximation approaches to perform the fitting of all experimental curves plotted.

The fitting model built from the equations consists in 5 unknown parameters $R_{ct}, R_{W0}, C_{dl}, \alpha$ and $\delta$. The electrolyte resistance is expected as fixe value as it is depending only on the electrolyte which is always the same (3 mM Ferri/Ferrocyanide in 0.5 M NaCl) [3]. In addition, the diffusion coefficient $D = 6{,}1.10^{-6}\mathrm{cm^2/s}$ was also fixed as the average value between the value of the coefficient diffusion of the Ferricyanide $D_{Ox} = 5{,}6.10^{-6}\mathrm{cm^2/s}$ and the Ferrocyanide $D_{Red} = 6{,}6.10^{-6}\mathrm{cm^2/s}$.

As the volumetric flow, $Q$, in the chip is a variable parameter depending to our experiments, it is interesting to keep the local diffusion layer thickness $\delta$ as a variable for in the model (**Figure 1D**). Indeed, the theoretical expression of $\delta$ is implemented in the code PyEIS with the following equation, as follows

$$\delta = 3^{\frac{2}{3}}\,\Gamma\left(\frac{4}{3}\right)\left(\frac{D.x_e}{S}\right)^{\frac{1}{3}} \quad (6)$$

where $S$ is the wall velocity gradient defined, as follows

$S = \frac{6.Q}{d.h^2}$,

$\Gamma$ the gamma-euler function, $x_e, d$ and $h$ the dimensions of the channel and the working electrode as depicted in **Figure 1**.
The fitting calculations have been performed with a least-squares fitting procedure [4] with a "modulus" weighting function calculated, as follows,

$\frac{1}{\sqrt{Re^2+Im^2}}$.

The work done in PyEIS mainly consisted in implementing the equation (4) in the code with the other pre-existing models, and adapting each function the module to be able to plot, simulate and fit this modified-Randles model.

As discussed previously, the resulting fitting function includes 5 parameters, the charge transfer resistance $R_{ct}$ (Ω), the Impedance of diffusion at a high frequency regime, $R_d$ (Ω), the CPE coefficient n, the non-ideal capacitance $Q$ (F), and the diffusion layer thickness δ (cm).

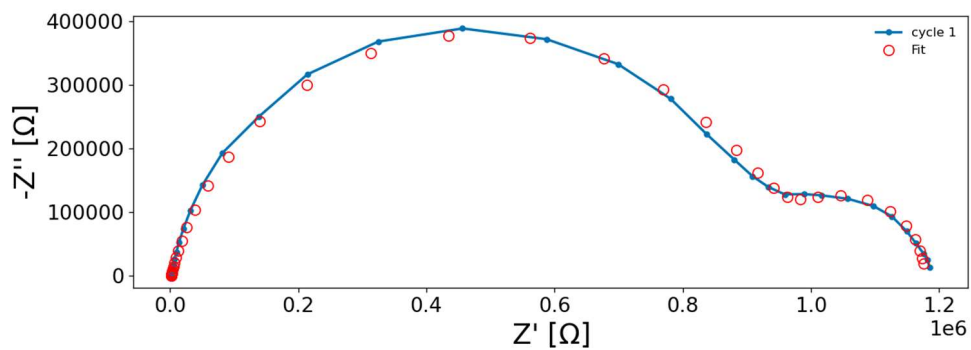

**Figure 2**: Example of experimental data and its corresponding fitting procedure. This example is referring to an experiment performed with Q=0.01µL/s, the minimum working flow that can be set using the model of neMESYS syringe pump available. (A) Nyquist EIS plot, experimental data (blue curve) and fitting model (red circle).

The function returns the fitted graph and the corresponding fitting results. The main drawback of this study, is the huge value of chi-square that is a problem still unsolved. Further works has to be done, especially on the weight coefficient (modulus) for the least square fitting method, that seems to be not appropriate for our model. The small incertitude values and the reported correlations are therefore comforting that the results aren't irrelevant.

## 4. Discussion

The limits of our model have been tested with fitting the EIS plot for several values of the volumetric flow, from 0.01 to 1 µL/s. The experiment is performed for a 30µm-width bare gold microelectrode.

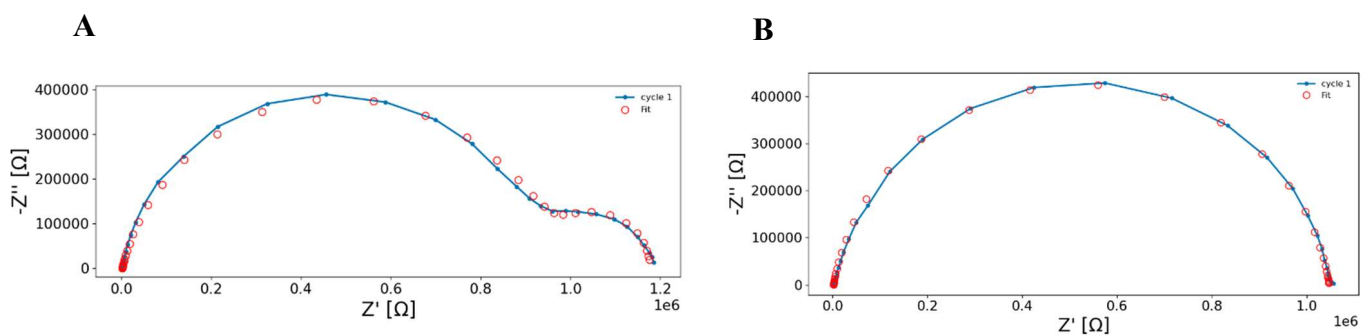

**Figure 3**: EIS plot on the same microfluidic channel with 3 mM of ferri/ferrocyanide diluted in 0.5 NaCl, for different flow rate: 0.01 µL/s (A) and 1 µL/s (B).

| $Q$ | $R_{ct}$ | $R_d$ | $Q$ | $n$ | $\delta$ |
|---|---|---|---|---|---|
| (µL/s) | (kΩm) | (kΩm) | nF | - | (µm) |
| **0.01** | 939.1 | 236.6 | 4.05 | 0.86 | 13.8 |
| **0.05** | 932.3 | 150.1 | 3.96 | 0.86 | 7.99 |
| **0.08** | 990.5 | 150.1 | 4.36 | 0.86 | 6.81 |
| **0.1** | 1057.9 | 150.0 | 3.76 | 0.88 | 5.54 |
| **0.2** | 1077.7 | 101.4 | 4.10 | 0.87 | 4.10 |
| **0.5** | 1050 | 100.2 | 4.05 | 0.88 | 2.40 |
| **0.8** | 989.4 | 124.2 | 3.82 | 0.89 | 2.09 |
| 1 | 930.9 | 114.0 | 4.41 | 0.89 | 2.25 |

**Table 1.** Evolution of parameters fitted values for different flow rates Q from 0.01 µL/s to 1 µL/s in the the same microchannel (with 3 mM of ferri/ferrocyanide diluted in 0.5 NaCl)

In the **Figure 3**, two EIS plot performed on the same chip with a flow rate equal to 0.01 µL/s (layout A) and 1 µL/s (layout B) are compared. A net change of shape of the EIS observed. The second loop, characteristic of the diffusion phenomena is not visible anymore for the highest value of the volumetric flow. Which is a meaningful result, as it confirms the fact that the diffusive transport is dominated by the convective transport for high value of volumetric flow. Another interesting phenomenon is related to the total impedance, and more precisely the real part of the impedance (*x*-axis) which is also slightly decreasing from 0.01 to 1 µl/s, which is confirming an increase of the current, as expected in channel microelectrode hydrodynamics. We report in **Table 2** the fitting results for the found parameter $\delta$ with confronting it to its theoretical values calculated using the equation (6) of the previous section, for several values of the volumetric flow, *Q*.

| $Q$ (µL/s) | $\delta$ theoretical (µm) | $\delta$ fitting (µm) |
|---|---|---|
| **0.01** | 11.4 | 13.8 |
| **0.05** | 6.64 | 7.99 |
| **0.08** | 5.68 | 6.81 |
| **0.1** | 5.27 | 5.54 |
| **0.2** | 4.19 | 4.10 |
| **0.5** | 3.08 | 2.40 |
| **0.8** | 2.64 | 2.09 |
| **1** | 2.45 | 2.25 |

**Table 2.** Evolution and comparison between theoretical and fitted values for $\delta$ with the microfluidic flow Q in the microchannel.

As expected, the diffusion layer thickness is decreasing when $Q$ is increasing. The decrease is though much more important from 0.01 to 0.2 µL/s and slows down strongly for the three-last value of flow (0.5, 0.8 and 1 µL/s). The expected decrease tendency is not anymore ensured for those values as we observe an increase from 0.8 and 1 µL/s.
It can therefore be guessed that the limit of our model is attained around those values of flow. As predicted experimentally by the previous work of MC-Horny, as the best value was set to $Q$=0.5 µL/s as the reference flow value for all the experiments.

**Conclusion**
The experiments consist to test several values of the flow rate, from 0.01 to 1 µL/s and then to simulate the EIS plots with the existing models for diffusion-convective impedance equation. Currently, the EIS Nyquist plots highlight two time constants that are traditionally observed for reactive impedance mixed with diffusion impedance due to the redox reaction-diffusion at microelectrode which is often modelled with the Randles equivalent circuit. The obtained results after implementing the diffusion-convective equation in the PyEIS code permits to adapt each function to be able to plot, simulate and fit a specific modified Randles for channel-electrode. The modified Randles was used for preliminary tests for optimization of the parameters of flow in the microfluidic channel by comparing the diffusion layer thickness as a function with the volumetric flow.

**Acknowledgments**
The authors would like to thank the DIMELEC project ANR-19-CE09-0016 for funding and RENATECH clean room facilities at C2N, Palaiseau, France.